\documentclass[12pt]{article}
\usepackage{graphicx}
\usepackage{amsmath}
\usepackage{url}
\pdfoutput=1

\newcommand\pubnumber{FERMILAB-CONF-14-537-ND}
\newcommand\pubdate{\today}

\def\indiana{Indiana University; Bloomington, IN 47405}
\def\support{\footnote{Author list at \url{http://www-boone.fnal.gov/collaboration}}}

\def\Title#1{\begin{center} {\Large #1 } \end{center}}
\def\Author#1{\begin{center}{ \sc #1} \end{center}}
\def\Address#1{\begin{center}{ \it #1} \end{center}}

\newcommand\pubblock{\rightline{\begin{tabular}{l} \pubnumber\\
         \pubdate  \end{tabular}}}
\newenvironment{Abstract}{\begin{quotation}  }{\end{quotation}}
\newenvironment{Presented}{\begin{quotation} \begin{center} 
             PRESENTED AT\end{center}\bigskip 
      \begin{center}\begin{large}}{\end{large}\end{center} \end{quotation}}

\def\beq{\begin{equation}}
\def\eeq#1{\label{#1}\end{equation}}
\def\eeqn{\end{equation}}

\def\beqa{\begin{eqnarray}}
\def\eeqa#1{\label{#1}\end{eqnarray}}
\def\eeqan{\end{eqnarray}}

\let\bar=\overbar

\def\L{{\cal L}}

\def\Dslash{\not{\hbox{\kern-4pt $D$}}}
\def\dslash{\not{\hbox{\kern-2pt $\del$}}}

\def\msb{{\bar{\ssstyle M \kern -1pt S}}}

\RequirePackage{xspace}

\def\MB        {MiniBooNE\xspace}
\def\DMS       {DM\xspace}

\def\SMS       {SM\xspace}
\def\SML       {Standard Model\xspace}

\def\unsim     {{\mathord{\sim}}}

\def\proton    {\ensuremath{p}\xspace}

\def\electron  {\ensuremath{e}\xspace}
\def\nubar     {\ensuremath{\overline{\nu}}\xspace}

\def\numu      {\ensuremath{\nu_\mu}\xspace}
\def\DMP       {\ensuremath{\chi}\xspace}

\def\DMV       {\ensuremath{V}\xspace}
\def\mDMP      {\ensuremath{m_\DMP}\xspace}
\def\mDMV      {\ensuremath{m_\DMV}\xspace}

\def\m         {\ensuremath{\rm \,m}\xspace}

\def\gev       {\ensuremath{\mathrm{\,Ge\kern -0.1em V}}\xspace}
\def\mev       {\ensuremath{\mathrm{\,Me\kern -0.1em V}}\xspace}
\def\kev       {\ensuremath{\mathrm{\,ke\kern -0.1em V}}\xspace}
\def\gevc      {\ensuremath{{\mathrm{\,Ge\kern -0.1em V\!/}c}}\xspace}

\def\mmev      {\ensuremath{{\mathrm{\,m\,Me\kern -0.1em V^{-1}}}\xspace}}

\begin{document}
\begin{titlepage}
\pubblock

\vfill
\Title{Accelerator-Produced Dark Matter Search using \MB}
\vfill
\Author{R. T. Thornton}
\Address{\indiana}
\Author{MiniBooNE-DM collaboration\support}
\vfill
\begin{Abstract}
  Cosmology observations indicate that our universe is composed of 25\% dark matter (\DMS), yet we know little about its microscopic properties. Whereas the gravitational interaction of \DMS is well understood, its interaction with the \SML is not. Direct detection experiments, the current standard, have a nuclear recoil interaction, low-mass sensitivity edge of order 1\gev. To detect \DMS with mass below 1\gev, either the sensitivity of the experiments needs to be improved or use of accelerators producing boosted low-mass \DMS are needed. Using neutrino detectors to search for low-mass \DMS is logical due to the similarity of the \DMS and \(\nu\) signatures in the detector. The \MB experiment, located at Fermilab on the Booster Neutrino Beamline, has produced the world's largest collection of \(\nu\) and \(\nubar\) samples and is already well understood, making it desirable to search for accelerator-produced boosted low-mass \DMS. A search for \DMS produced by 8.9\gevc protons hitting a steel beamdump has finished, collecting \(1.86\times10^{20}~\mathrm{POT}\). Analysis techniques along with predicted sensitivity will be presented.
\end{Abstract}
\vfill
\begin{Presented}
XXXIV Physics in Collision Symposium \\
Bloomington, Indiana,  September 16--20, 2014
\end{Presented}
\vfill
\end{titlepage}
\def\thefootnote{\fnsymbol{footnote}}
\setcounter{footnote}{0}

\section{Introduction}
Since the first proposal of dark matter (\DMS) by Zwicky~\cite{Zwicky} there has been overwhelming cosmological evidence of its existence~\cite{Bertone:2004pz}. To determine how \DMS interacts with the \SML of particle physics (\SMS), extensive searches for \DMS have been done. These searches include looking for a nuclear recoil signature in direct detection experiments. Direct detection experiments have a \DMS mass threshold around 1\gev. Boosting the \DMS will help probe masses less than a \gev. A study for \DMS masses less than a \gev is necessary to understand \DMS in general.

The \MB detector~\cite{AguilarArevalo:2008qa} is a \(\nu\) detector filled with 800~tons of mineral oil acting as a Cherenkov/scintillator detector. The light is read out by 1280 inner and 240 veto PMTs. \MB has been taking data for over 10 years obtaining \(6.5\times10^{20}~\mathrm{POT}\) (\(11.3\times10^{20}~\mathrm{POT}\)) in \(\nu\)(\(\nubar\)) mode. Publications include results for \(\nu\)-~\cite{AguilarArevalo:2010cx} and \(\nubar\)-~\cite{Aguilar-Arevalo:2013nkf} nucleon neutral current elastic scatting (NCel). Because it is such a well-understood detector, it is a good candidate to search for production of a vector mediator produced in a beamstop decaying into a \DMS particle which is detected in the \MB detector 490\m away. The beam is composed of 8.9\gevc protons.

\section{Benchmark Theoretical Model}
\begin{figure}[htbp]
  \begin{minipage}{0.49\textwidth}
    \includegraphics[width=\textwidth]{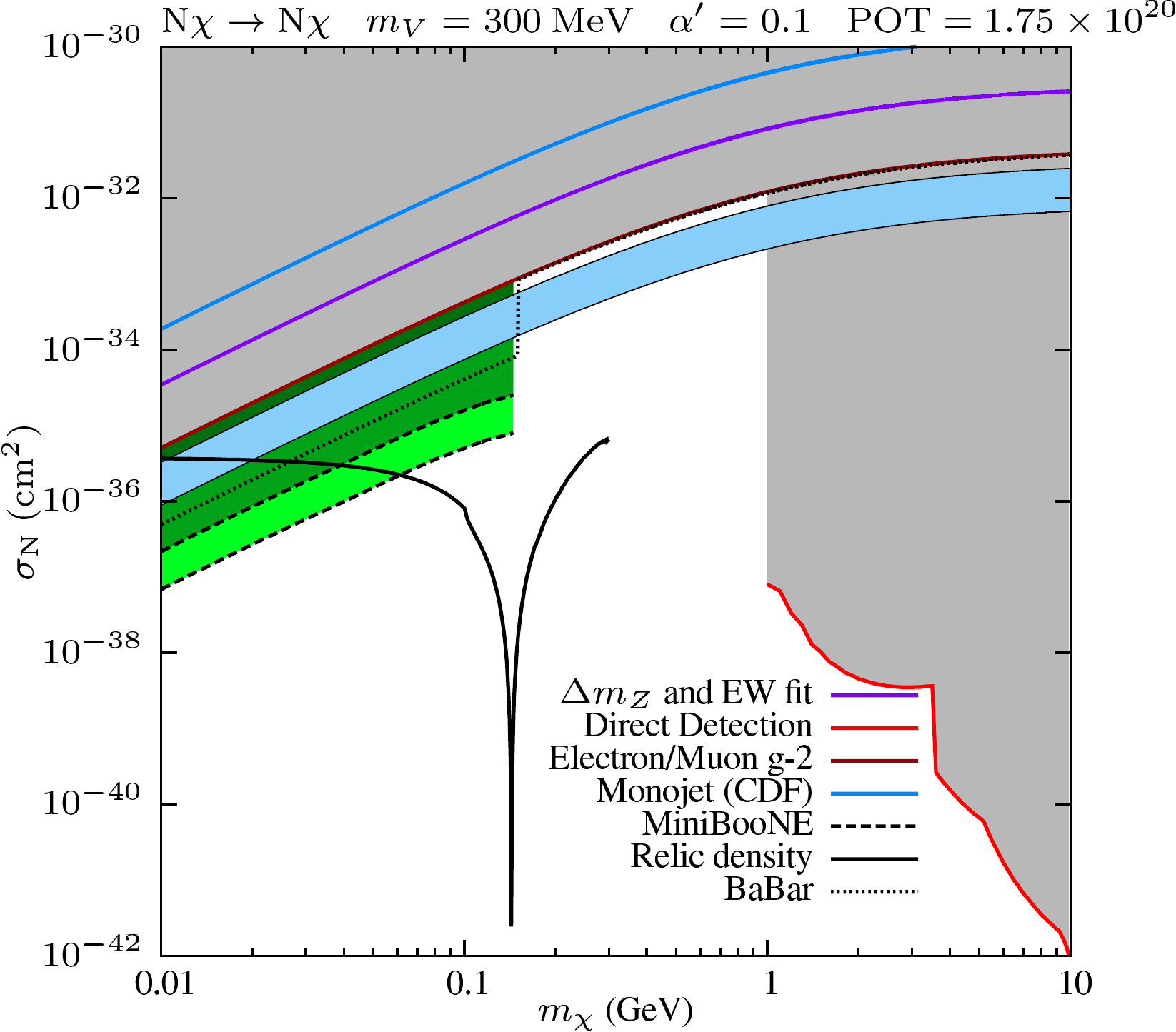}
  \end{minipage}\hspace*{0.01\textwidth}%
  \begin{minipage}{0.49\textwidth}
    \includegraphics[width=\textwidth]{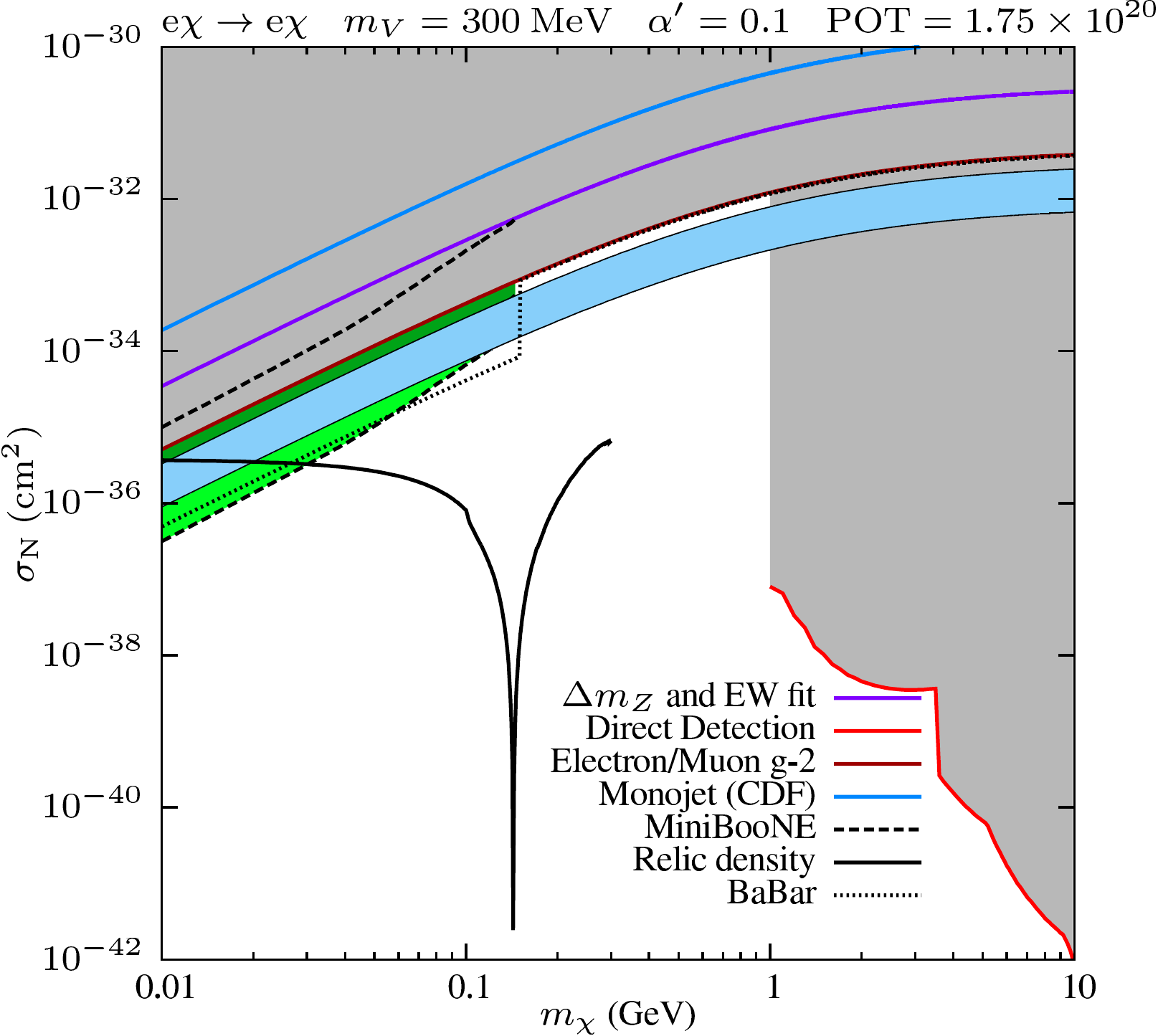}
  \end{minipage}
  \caption{Regions of \DMS-nucleon scattering cross section (corresponding to non-relativistic spin-independent coherent scattering on nuclei) vs. \DMS mass. In this plot we have fixed \DMV = 300\mev and \(\alpha^\prime = 0.01\). Constraints are shown from monojet searches (\(\proton\proton\rightarrow j+inv\))~\cite{Shoemaker:2011vi}, excessive contributions to (g-2)\(_\mu\)~\cite{Pospelov:2008zw}, precision electroweak measurements~\cite{Hook:2010tw}, a monophoton search (\(\electron^{+}\electron^{-}\rightarrow\gamma+inv\))~\cite{Aubert:2008as} (labeled BaBar), and low-mass limits from \DMS direct detection experiments, DAMIC~\cite{Barreto:2011zu} (1-3\gev), CDMSlite~\cite{Agnese:2013jaa} (3-5\gev) and XENON10~\cite{Angle:2011th} (5-10\gev). Note that a slightly stronger exclusion contour to XENON10 has recently been obtained by LUX~\cite{Akerib:2013tjd}. The light blue band represents the region where the current \(\unsim3\sigma\) discrepancy in (g-2)\(_\mu\) is alleviated by the 1-loop contribution from the vector mediator~\cite{Pospelov:2008zw}. The solid black line indicates where the relic density of the \DMS matches observations-the structure in this contour is due to s-channel \(\DMV^*\) resonant enhancement in the \DMS annihilation cross section for \(\mDMP \sim \mDMV/2\). For \(\mDMP > \mDMV/2\), new annihilation channels open up and this relation is modified. The left panel shows regions where we expect 1--10 (light green), 10--1000 (green), and more than 1000 (dark green) elastic scattering events off nucleons in the \MB detector with \(1.75\times10^{20}~\mathrm{POT}\). The right panel shows the same for elastic scattering off electrons.}
  \label{fig:predicted_plots}
\end{figure}

\begin{figure}[htbp]
    \begin{minipage}{0.49\textwidth}
      \includegraphics[width=\textwidth]{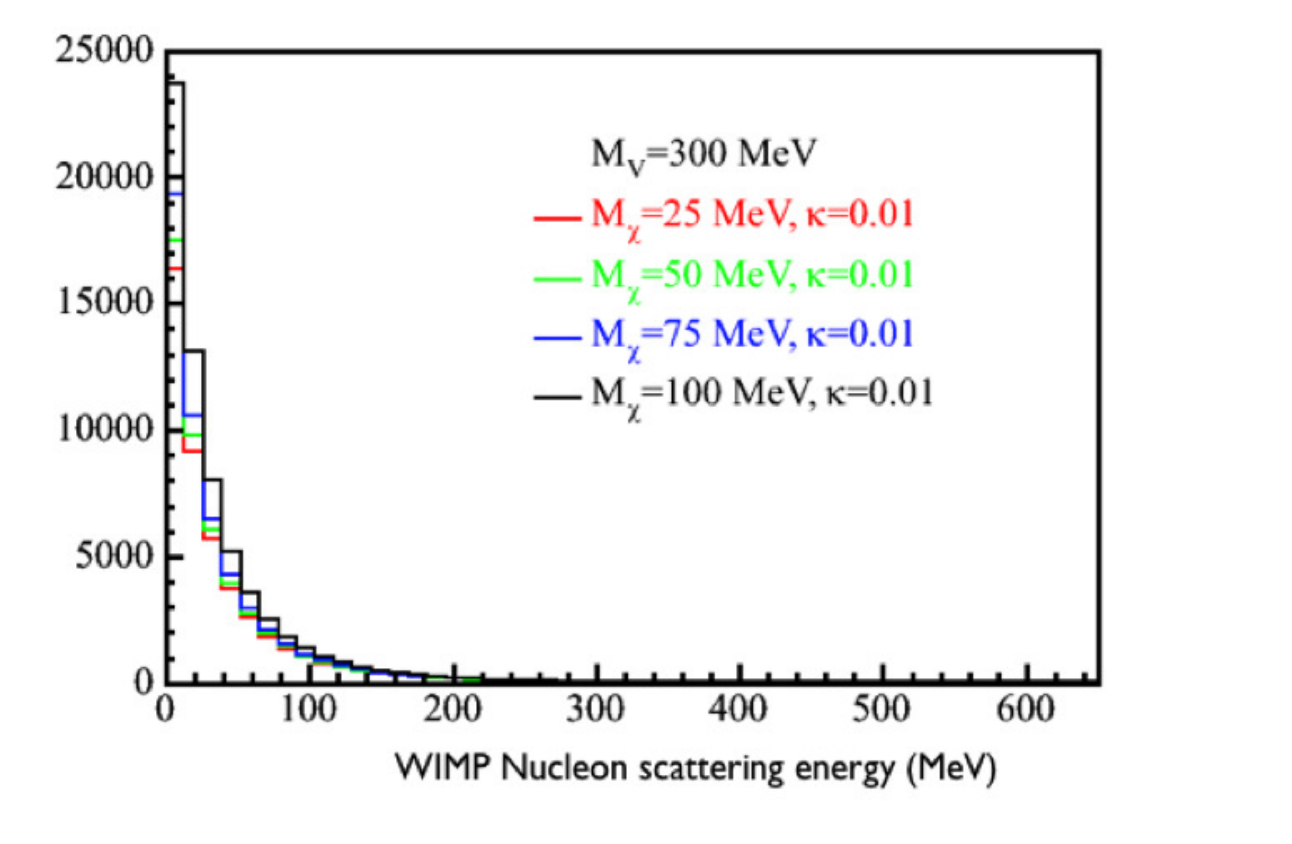}
    \end{minipage}\hspace*{0.01\textwidth}%
    \begin{minipage}{0.49\textwidth}
      \includegraphics[width=\textwidth]{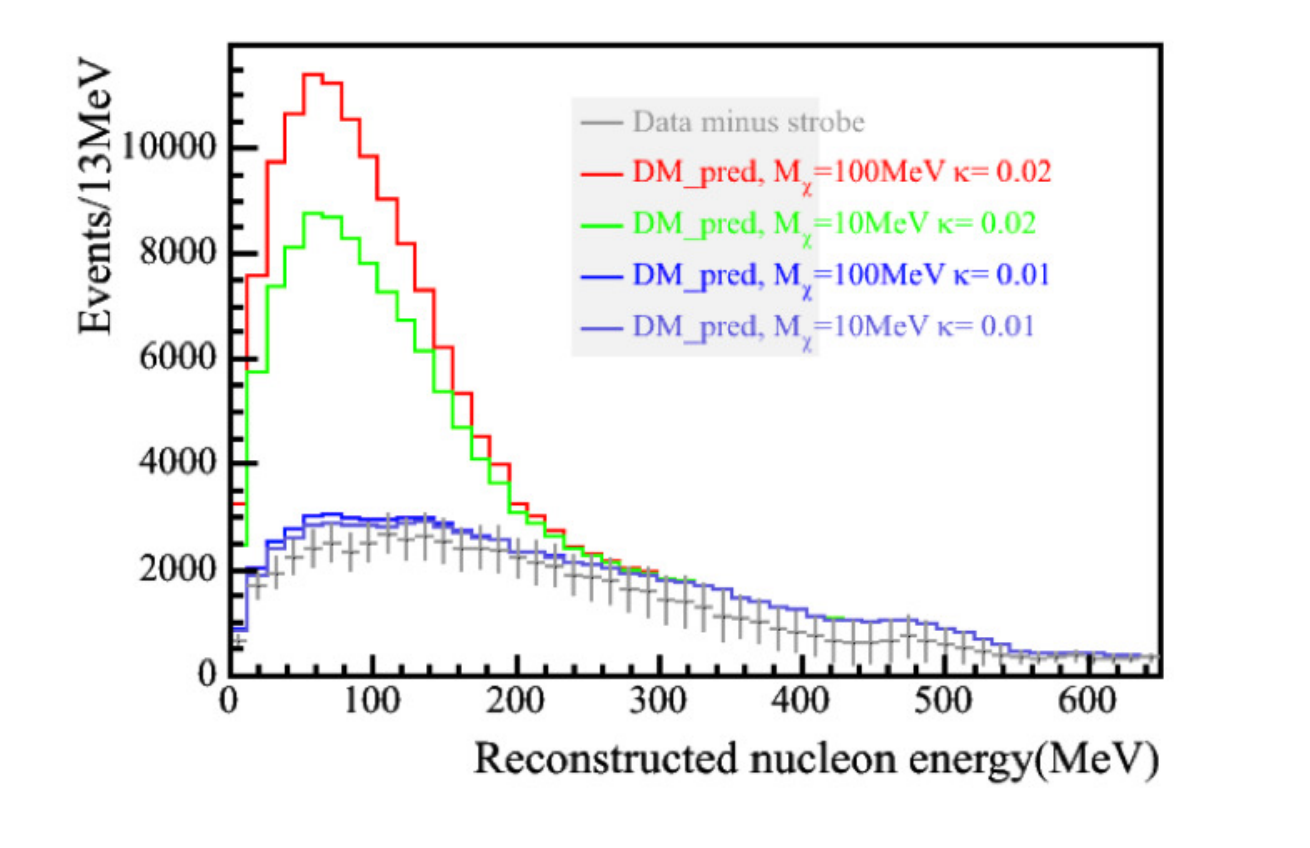}
    \end{minipage}
      \caption{Nucleon kinetic energy from \DMS scattering for various model parameters (left). Reconstructed nucleon kinetic energy for both \DMS and neutrino scattering after corrected for detector efficiency (right).}
      \label{fig:signal}
\end{figure}

The benchmark model is a vector portal model of light \DMS. It was initially developed by Boehm and Fayet~\cite{Boehm:2003hm} in order to solve the 511\kev spectrum line seen from the galactic bulge. The theory consists of \DMV a vector mediator with a mass \mDMV less than 1\gev and the \DMS particle \DMP with a mass \mDMP. The following is added to the \SMS Lagrangian~\cite{Batell:2014mga}:
\begin{equation}
  \begin{split}
    \L &=\L_\DMP-\frac{1}{4}F_{\mu\nu}^\prime F^{\prime\mu\nu}+\frac{1}{2}\mDMV^2\DMV_{\mu}\DMV^{\mu}-\frac{\kappa}{2}F_{\mu\nu}F^{\prime\mu\nu}\\
    \L_\DMP &= 
    \left\{
      \begin{array}{l l}
        \imath\overline{\DMP}\left(\partial_\mu-\imath g^\prime \DMV_{\mu}\right)\DMP - \mDMP\overline{\DMP}\DMP & \quad\text{Dirac fermion \DMS}\\
        \left|\left(\partial_\mu-\imath g^\prime \DMV_{\mu}\right)\DMP\right|^2 - \mDMP^2\left|\DMP\right|^2 & \quad\text{Complex scalar \DMS}
      \end{array}
      \right.
    \end{split}
  \end{equation}
  which leads to four new free parameters, (i) \mDMP, (ii) \mDMV, (iii) kinetic mixing angle between \DMS and \SMS, \(\kappa\), and (iv) a coupling constant for \DMS and \DMP, \(\alpha^\prime=g^\prime/4\pi\). When \(\DMV < 2\mDMP\) the decay is practically visible, into leptons or photons, and invisible otherwise. The production of \DMV can be produced in beamlines from either through Bremsstrahlung or decay of neutral mesons. References~\cite{Pospelov:2007mp} apply the benchmark theory to several neutrino experiments, predicting the experiments sensitivity. \MB has the best sensitivity, in the interesting parameter regions, when the decay is invisible (Fig.~\ref{fig:predicted_plots}), where the signature of \DMP is either NCel off a nucleon or an electron.
Due to detector efficiency and background, \MB is most sensitive when the reconstruction nuclear recoil is between 40 and 250\mev (Fig.~\ref{fig:signal}).

\section{Beamdump Mode}
\begin{figure}[htbp]
  \begin{minipage}{0.25\textwidth}
    \includegraphics[width=\textwidth]{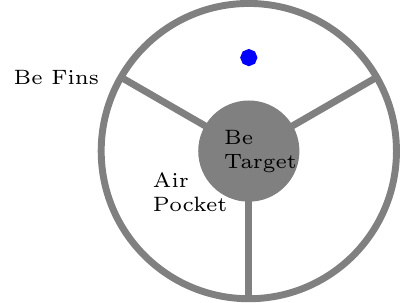}
  \end{minipage}\hspace*{0.01\textwidth}%
  \begin{minipage}{0.749\textwidth}
    \includegraphics[width=\textwidth]{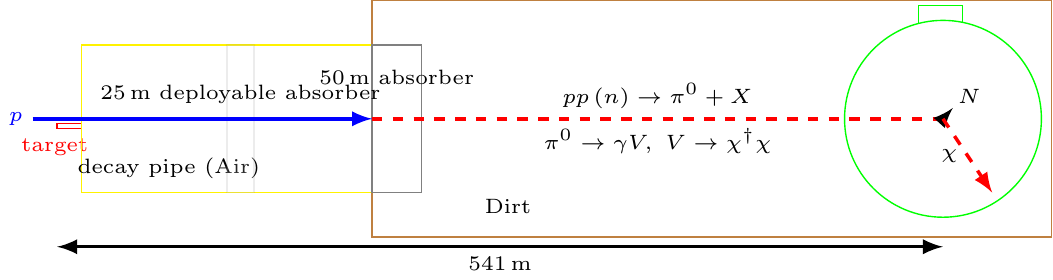}
  \end{minipage}
  \caption{Schematic of beam spot when running in beamdump mode (left). Cartoon of \DMS production in beamdump mode (right).}
  \label{fig:beamOffTarget}
\end{figure}

Running \MB in beamdump mode reduces \(\nu\)-induced events, which are a background to a \DMS measurement, by having the beam hit a steel beamdump 50\m downstream instead of hitting the Be target (Fig.~\ref{fig:beamOffTarget}).
A reduction factor of \(\unsim\)40 from beam on-target with the horn in neutrino mode to beamdump mode is seen by comparing the \(\numu\)CCQE analyses. \(\numu\)CCQE is also used to determine if an extra scale factor to the simulations is needed.

\section{Semi-Blind Preliminary Results}
\begin{figure}[htbp]
  \begin{minipage}{0.33\textwidth}
    \includegraphics[width=\textwidth]{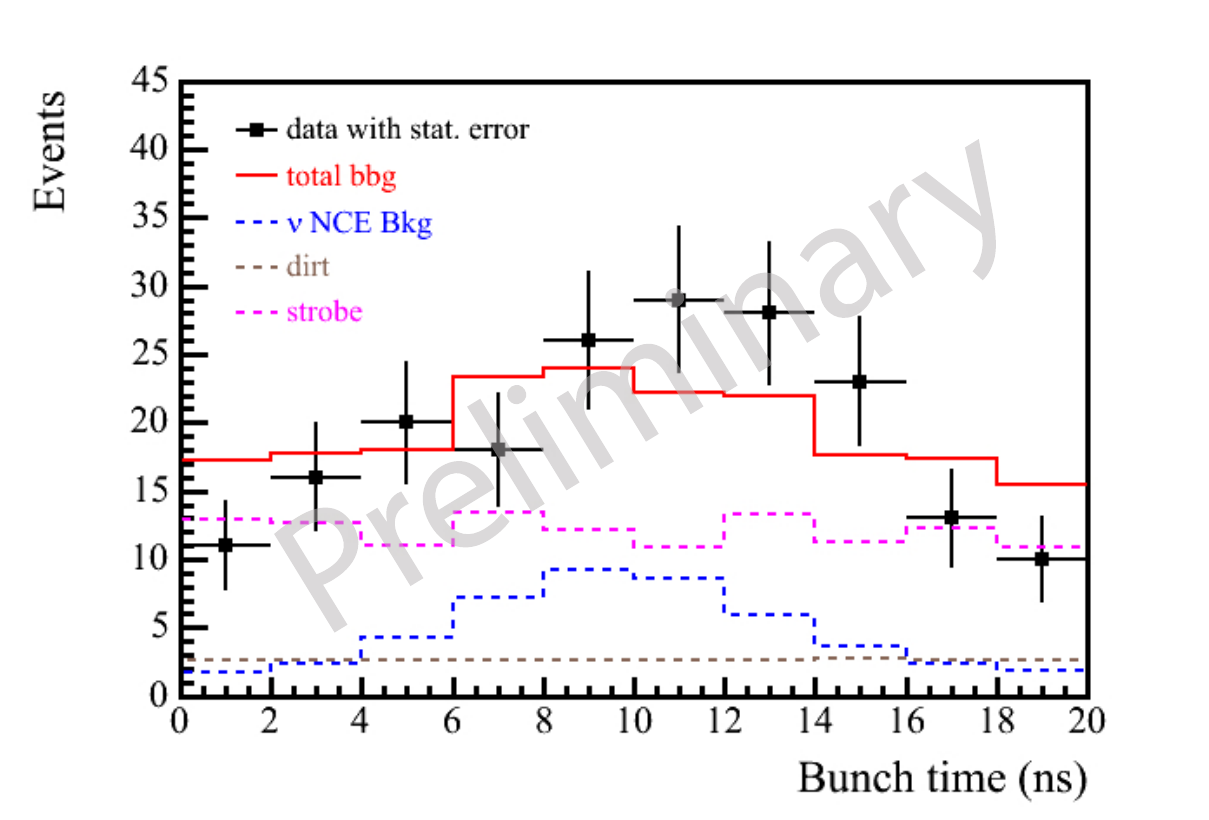}
  \end{minipage}\hspace*{0.005\textwidth}%
  \begin{minipage}{0.33\textwidth}
    \includegraphics[width=\textwidth]{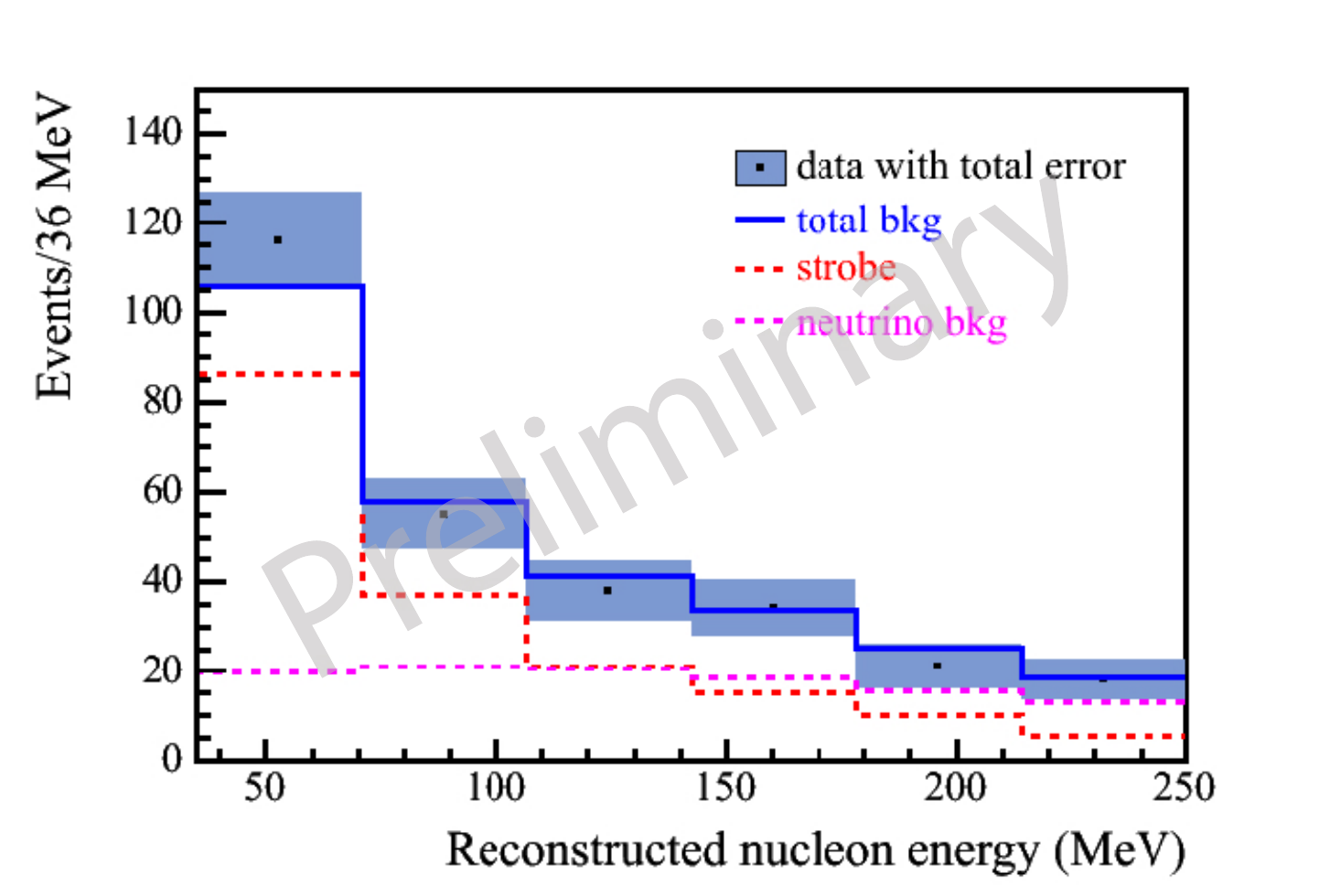}
  \end{minipage}\hspace*{0.005\textwidth}%
  \begin{minipage}{0.33\textwidth}
    \centering
    \footnotesize
    \begin{tabular}{|r|l|}
      \hline
      & \# events\\
      \hline
      Cosmics & 177 \\
      Total \(\nu\) bkg & 107.8 \\
      Total Bkg & \(284.8\pm18\) (sys.) \\
      Data & \(284\pm17\) (stat)\\
      \hline
    \end{tabular}
    \normalsize
  \end{minipage}
  \caption{Time of events inside the proton bunch (left). Energy distributions for data and background (middle). Table showing integral of the energy distributions (right).}
  \label{fig:results}
\end{figure}

A semi-blind analysis is being done. The first \(3.2\times10^{19}~\mathrm{POT}\) of total \(1.86\times10^{20}~\mathrm{POT}\) is used to adjust cuts and understand predictions from simulations. For the nucleon analysis, two observables will be used to compare to theory, nuclear recoil energy and \DMS time of flight. Using both observables will reduce backgrounds and give better information about the \DMS mass; there is a correlation between \DMS time of flight and mass. We use the same event selection as~\cite{Aguilar-Arevalo:2013nkf} except the nuclear recoil energy is now 35--250\mev. With these cuts, we see the results in Fig.~\ref{fig:results}

\section{Conclusion}
\MB has finished a beamdump run, accumulating a total of \(1.86\times10^{20}~\mathrm{POT}\). The motivation of this run was to search for \DMS particles with a mass less than 1\gev. In the light \DMS model, \MB has been shown to be able to add to the exclusion plots. Doing a semi-blind analysis, preliminary results on \DMP-nucleon NCel shows a consistency with null. Final results expected fall 2015.



\begin{thebibliography}{99}


\bibitem{Zwicky}
F. Zwicky Helv. Phys. Acta 6 (1933) 110.

\bibitem{Bertone:2004pz} 
G.~Bertone, D.~Hooper and J.~Silk, Phys.\ Rept.\  {\bf 405}, 279 (2005) [hep-ph/0404175].

\bibitem{AguilarArevalo:2008qa} 
A.~A.~Aguilar-Arevalo {\it et al.}  [\MB Collaboration], Nucl.\ Instrum.\ Meth.\ A {\bf 599}, 28 (2009) [arXiv:0806.4201 [hep-ex]].

\bibitem{AguilarArevalo:2010cx} 
A.~A.~Aguilar-Arevalo {\it et al.}  [\MB Collaboration], Phys.\ Rev.\ D {\bf 82}, 092005 (2010) [arXiv:1007.4730 [hep-ex]].

\bibitem{Aguilar-Arevalo:2013nkf} 
A.~A.~Aguilar-Arevalo {\it et al.}  [\MB Collaboration], arXiv:1309.7257 [hep-ex].

\bibitem{Boehm:2003hm} 
C.~Boehm and P.~Fayet, Nucl.\ Phys.\ B {\bf 683}, 219 (2004) [hep-ph/0305261].

\bibitem{Batell:2014mga} 
B.~Batell, R.~Essig and Z.~Surujon, Phys.\ Rev.\ Lett.\  {\bf 113}, no. 17, 171802 (2014) [arXiv:1406.2698 [hep-ph]].


\bibitem{Pospelov:2007mp} 
M.~Pospelov, A.~Ritz and M.~B.~Voloshin, Phys.\ Lett.\ B {\bf 662}, 53 (2008) [arXiv:0711.4866 [hep-ph]];
B.~Batell, M.~Pospelov and A.~Ritz, Phys.\ Rev.\ D {\bf 80}, 095024 (2009) [arXiv:0906.5614 [hep-ph]];
P.~deNiverville, M.~Pospelov and A.~Ritz, Phys.\ Rev.\ D {\bf 84}, 075020 (2011) [arXiv:1107.4580 [hep-ph]];
P.~deNiverville, D.~McKeen and A.~Ritz, Phys.\ Rev.\ D {\bf 86}, 035022 (2012) [arXiv:1205.3499 [hep-ph]].

\bibitem{Shoemaker:2011vi} 
I.~M.~Shoemaker and L.~Vecchi, Phys.\ Rev.\ D {\bf 86}, 015023 (2012) [arXiv:1112.5457 [hep-ph]].

\bibitem{Pospelov:2008zw} 
M.~Pospelov, Phys.\ Rev.\ D {\bf 80}, 095002 (2009) [arXiv:0811.1030 [hep-ph]].

\bibitem{Hook:2010tw} 
A.~Hook, E.~Izaguirre and J.~G.~Wacker, Adv.\ High Energy Phys.\  {\bf 2011}, 859762 (2011) [arXiv:1006.0973 [hep-ph]].

\bibitem{Aubert:2008as} 
B.~Aubert {\it et al.}  [BaBar Collaboration], arXiv:0808.0017 [hep-ex];
E.~Izaguirre, G.~Krnjaic, P.~Schuster and N.~Toro, Phys.\ Rev.\ D {\bf 88}, 114015 (2013) [arXiv:1307.6554 [hep-ph]];
R.~Essig, J.~Mardon, M.~Papucci, T.~Volansky and Y.~M.~Zhong, JHEP {\bf 1311}, 167 (2013) [arXiv:1309.5084 [hep-ph]].

\bibitem{Barreto:2011zu} 
J.~Barreto {\it et al.}  [DAMIC Collaboration], Phys.\ Lett.\ B {\bf 711}, 264 (2012) [arXiv:1105.5191 [astro-ph.IM]].

\bibitem{Agnese:2013jaa} 
R.~Agnese {\it et al.}  [SuperCDMS Collaboration], Phys.\ Rev.\ Lett.\  {\bf 112}, no. 4, 041302 (2014) [arXiv:1309.3259 [physics.ins-det]].

\bibitem{Angle:2011th} 
J.~Angle {\it et al.}  [XENON10 Collaboration], Phys.\ Rev.\ Lett.\  {\bf 107}, 051301 (2011) [Erratum-ibid.\  {\bf 110}, 249901 (2013)] [arXiv:1104.3088 [astro-ph.CO]].

\bibitem{Akerib:2013tjd} 
D.~S.~Akerib {\it et al.}  [LUX Collaboration], Phys.\ Rev.\ Lett.\  {\bf 112}, no. 9, 091303 (2014) [arXiv:1310.8214 [astro-ph.CO]].

\end{thebibliography}
\end{document}